\documentclass[aps,pra,twocolumn,showpacs,floatfix]{revtex4}

\usepackage{graphicx}
\usepackage{graphics}
\usepackage{amsmath}

\begin{document}

\title{Semiclassical limits to the linewidth of an atom laser}
\author{Mattias Johnsson}
\author{Simon Haine}
\author{Joseph Hope}
\author{Nick Robins}
\author{Cristina Figl}
\author{Matthew Jeppesen}
\author{Julien Dugu\'{e}}
\author{John Close}

\affiliation{Australian Centre for Quantum-Atom Optics, Physics Department, The Australian National University, Canberra, 0200, Australia.}
\email{mattias.johnsson@anu.edu.au}
\homepage{http://www.acqao.org}

\begin{abstract}
We investigate the linewidth of a quasi-continuous atom laser within a semiclassical framework.  In the high flux regime, the lasing mode can exhibit a number of undesirable features such as density fluctuations.  We show that the output therefore has a complicated structure that can be somewhat simplified using Raman outcoupling methods and energy-momentum selection rules.  In the weak outcoupling limit, we find that the linewidth of an atom laser is instantaneously Fourier limited, but, due to the energy `chirp' associated with the draining of a condensate, the long-term linewidth of an atom laser is equivalent to the chemical potential of the condensate source.  We show that correctly sweeping the outcoupling frequency can recover the Fourier-limited linewidth. \end{abstract}


\maketitle
\section{Introduction}

Optical lasers have found broad application in precision measurements that address questions both fundamental and applied in nature.  In many cases, we expect to be able to perform such experiments more effectively and to higher precision with atom interferometry \cite{kasevich, berman}.  Ultra-cold ensembles of thermal atoms are already utilized in
interferometric systems where they have been demonstrated to compete with the best inertial and gravitational measurement apparatus available \cite{kasevich,petersET1999,gustavsonET2000,mcguirkET2002,fattoriET2003}. For this reason, there is significant interest in the production of a coherent source of ultra-cold atoms from a Bose-Einstein condensate (BEC) for applications in precision measurement and metrology \cite{wichtET2001,cladeET2006}. A free-space BEC atom interferometer in the Mach-Zehnder configuration has been demonstrated to produce 100\% contrast at the output port \cite{toriiET2000}, and there are proposals to put atom interferometers into space \cite{lecoqET2006}.  Due to trap stability and mean-field effects, precision experiments will most likely need to be performed with the low density, untrapped atomic beam rather than in more dense atomic sources such as full Bose Einstein condensates \cite{lecoqET2006}.    There is also interest in producing and measuring non-classical quantum states of atomic beams \cite{reid,kheruntsyan2005,haine05}.  All of these proposed applications will require a spatially stable atom laser beam with good first order coherence.  

In this paper we investigate the key spatial properties of a quasi-continuous atom laser.  In the cases where the outcoupled atom laser is stable, we focus on the linewidth of the output spectrum as the key measure of the first-order coherence of the beam.  We begin by highlighting the particular importance of linewidth for dispersive fields such as atoms.  In Sec.~\ref{secModel} we introduce our model, and we calculate the properties of the output in various limits in Sec.~\ref{secLinewidth}.  In Sec.~\ref{secLinewidthReduction} we describe methods for reducing the atom laser linewidth.

\section{The importance of linewidth}
In a precision interferometric measurement made at the shot noise limit, all that is theoretically required of the wave source, whether it be a source of matter waves or light, is that it have high flux.  In principle, classical source fluctuations in frequency and phase can be removed through good interferometer design.  A long coherence length, equivalent to a spectrally narrow source, is not required if the path length difference in the interferometer is less than the coherence length.  In principle, mode matching on the output beam splitter of an interferometer can be performed as well on a complicated spatial mode as a simple one and a highly divergent  beam can be collimated with lenses.  In practice, however, if an interferometer is to operate at the shot noise limit, none of this is true. The shot noise limit for a high flux source is difficult to achieve, and it is essential to have a spectrally narrow, classically quiet, low divergence beam with the minimum transverse structure in both phase and amplitude. For these reasons, typical precision optical measurements use classically quiet lasers operating on the TEM$_{00}$ mode.  

\begin{figure}[htb]
\begin{center}
\includegraphics[width=7cm,height=2.8cm]{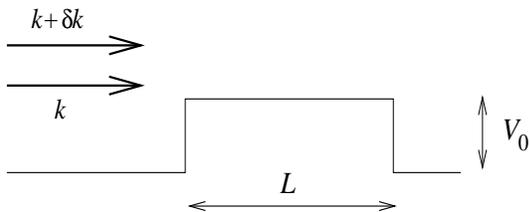}
\caption{Two matter waves with wave vectors $k$ and $k+\delta k$ are incident on a step potential of height $V_0$.} 
\label{figStepFunctionPotential}
\end{center}
\end{figure}

These points apply equally well to optical and atom lasers, suggesting that in the absence of a perfect experiment, a spectrally pure output beam is highly desirable. The difference is that unlike in an optical interferometer, the linewidth of a matter wave interferometer can still be critical even for a perfect experiment. To see this we consider the case of an equal path length Mach-Zehnder interferometer and an atomic beam that has an average momentum $\hbar k$ and a momentum spread of $\hbar \, \delta k$. We now introduce some disturbance in one of the arms, corresponding to the effect we wish to measure, and model it as a step function potential of width $L$ and height $V_0$. This is shown schematically in Figure \ref{figStepFunctionPotential}.

Assuming that $\delta k \ll k$, it is easily shown that due to the dispersive nature of the atomic beam a phase difference of
\begin{equation}
\Delta \phi = \frac{L \delta k}{2} \left[ \frac{\hbar k (\hbar^2 k^2 - 4 m V_0)} {(\hbar^2 k^2 - 2 m V_0)^{3/2}} - 1\right]
\label{eqStepPotentialPhaseDifference}
\end{equation}
builds up between the $k$ and $k+\delta k$ components of the incident beam. The phase difference $\Delta \phi$ represents an inherent uncertainty in the phase resolution of the interferometer, regardless of what measurement technique is used at the output ports. This uncertainty arises purely from the finite linewidth of the atomic beam, and will exist no matter how accurately the path lengths of the interferometer arms are matched.

Now consider the case for an optical interferometer, again using spectrally broad beams, and again with path lengths perfectly matched. If a potential is introduced into one of the arms, it corresponds to a change in the refractive index over that region. Provided this change is not dispersive, all the components in the beam will see the potential simply as an increase in path length --- crucially, the {\emph{same}} increase. Consequently the interferometer can be nulled and brought back to correct operation by physically readjusting the distance in one of the arms. As indicated in Eq.\,(\ref{eqStepPotentialPhaseDifference}), this readjustment is not possible in an atom interferometer.

In general, generating an atomic source with low linewidth also requires the beam to have well-controlled spatial properties in other respects, which makes it a good choice of metric for the first order coherence of atomic sources.  Applications that do not specifically benefit from the narrow linewidth of an atom laser will still tend to benefit from the associated controlled spatial mode.

\section{Model} \label{secModel}

In the most general terms, an atom laser requires coupling atoms out of a BEC into a coherent beam. The most common way to accomplish this is to use a state-changing outcoupling method, where the atomic species making up the BEC has at least two separate internal states --- a trapped state and an untrapped state. The BEC consists of atoms in the trapped state, which feel some confining potential that keeps them localized and Bose condensed. Some external perturbation is then applied to the BEC which flips a portion of the trapped atoms into the untrapped state where they no longer experience the confining potential and are free to leave the trap.

\begin{figure}[htb]
\begin{center}
\includegraphics[width=7cm,height=5cm]{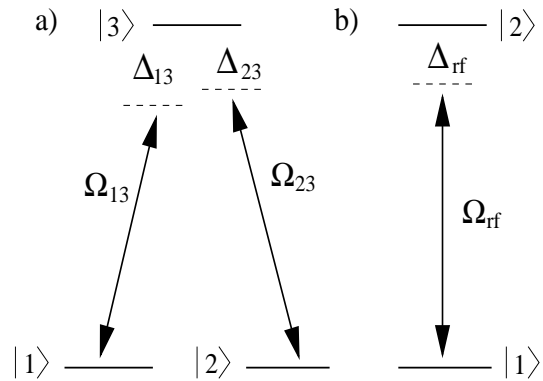}
\caption{An atom laser based on a) Raman outcoupling and b) rf outcoupling. In both cases trapped atoms in state $| 1 \rangle$ are transferred to an untrapped state $| 2 \rangle$ via electromagnetic fields. The $\Omega_{ij}$ represent the Rabi frequency of the applied fields and the $\Delta_{ij}$ represent detuning from resonance.} 
\label{figlevel_scheme_rf_and_raman}
\end{center}
\end{figure}

There are two common outcoupling methods, both of which require that the atoms of the BEC are in a specific angular momentum substate, for example $m_F =1$, meaning they can be confined by a magnetic trap. The two methods are shown schematically in Figure~\ref{figlevel_scheme_rf_and_raman}.

In the first outcoupling scheme, an external rf field is applied that flips the atoms into an $m_F = 0$ state that does not see the trapping magnetic field. Consequently the atoms fall from the trap under the influence of gravity, creating a semi-directed beam of coherent atoms: the atom laser. In the second scheme a Raman outcoupling method is used, where two optical fields transfer the state of the atoms from trapped ($m_F = 1$) to untrapped ($m_F = 0$) via a third intermediate level. The use of two optical fields means that a significant momentum kick can be imparted to the untrapped atoms as they leave the condensate, leading to an atom laser with superior properties such as a higher flux and higher brightness \cite{robinsET2006} and the ability to give the beam directionality \cite{hagleyET1999, robinsET2006}. In addition, the Raman scheme allows for the possibility of creating non-classical states of the beam \cite{haine05,haineET2006}.

Regardless of whether an rf or a Raman outcoupling scheme is used, the second-quantized Hamiltonian describing the system can be written most generally as
\begin{equation}
\hat{H} = \int \left( \hat{H}_{\mathrm{trap}} + \hat{H}_{\mathrm{beam}} + \hat{H}_{\mathrm{int}} \right) d^3{\mathbf{r}}
\label{eqGeneralHamiltonian}
\end{equation}
where $\hat{H}_{\mathrm{trap}}$ describes the atoms in the trap, $\hat{H}_{\mathrm {beam}}$ describes the atoms in the atom laser beam, and $\hat{H}_{\mathrm {int}}$ describes the outcoupling process. For the purposes of this paper we assume an isotropic harmonic trapping potential for the trapped atoms, which enables us to write the terms of Eq.\,(\ref{eqGeneralHamiltonian}) with greater specificity as
\begin{eqnarray}
\hat{H}_{\mathrm{trap}} &=& \hat{\Psi}^{\dagger}_t \left( -\frac{\hbar^2}{2m}\nabla^2 + \frac{1}{2} m \omega_t^2 r^2 + \frac{U_{tt}}{2} \hat{\Psi}^{\dagger}_t \hat{\Psi}_t \right) \hat{\Psi}_t \label{eqHtrap} \\
\hat{H}_{\mathrm{beam}} &=& \hat{\Psi}^{\dagger}_u \left( -\frac{\hbar^2}{2m}\nabla^2  - \hbar \delta + \frac{U_{uu}}{2} \hat{\Psi}^{\dagger}_u \hat{\Psi}_u \right)  \hat{\Psi}_u \label{eqHbeam} \\
\hat{H}_{\mathrm{int}} &=& - \hbar \left( \Omega ({\mathbf{r}}) \hat{\Psi}_u^{\dagger} \hat{\Psi}_t + \Omega^* ({\mathbf{r}}) \hat{\Psi}_t^{\dagger} \hat{\Psi}_u \right. \nonumber \\
&& \hspace{1cm} \left. + U_{tu}  \hat{\Psi}^{\dagger}_t \hat{\Psi}^{\dagger}_u \hat{\Psi}_u \hat{\Psi}_t \right) \label{eqHcoupling}
\end{eqnarray}
where $\hat{\Psi}_t({\mathbf{r}})$ and $\hat{\Psi}_u({\mathbf{r}})$ describe the trapped and untrapped matter fields respectively, $\omega_t$ is the harmonic trapping frequency, and the $U_{ij}$ are nonlinear potentials arising from atom-atom collisions. For rf outcoupling the detuning is given by $\delta = \Delta_{\mathrm{rf}}$ and for Raman outcoupling it is given by
\begin{equation}
\delta = \Delta_{23} - \Delta_{13} + \frac{|\Omega_{23}|^2}{\Delta_{13}} - \frac{|\Omega_{13}|^2}{\Delta_{13}}.
\end{equation}
The specifics of the outcoupling strength $\Omega ({\mathbf{r}})$ depend on the outcoupling method. In the rf outcoupling case $\Omega = \Omega_{\mathrm{rf}}$ and has no position dependence. With Raman outcoupling, however, $\Omega$ is position dependent and is given by
\begin{equation}
\Omega({\mathbf{r}}) = \frac{\Omega_{13}^* \Omega_{12}}{\Delta_{13}} e^{i{\mathbf{k}}_0 \cdot {\mathbf{r}}}
\end{equation}
where ${\mathbf{k}}_0 = {\mathbf{k}}_2 - {\mathbf{k}}_1$ is the momentum kick imparted to the outcoupled atoms from the two optical beams.

We assume that the atomic gas is sufficiently cold and dilute, so only binary collisions are relevant and the nonlinear potentials are defined by 
\begin{equation}
U_{ij} = 4 \pi \hbar^2 a_{ij}/m,
\label{eqNonlinearPotentialDefinition}
\end{equation}
where $a_{ij}$ is the $s$-wave scattering length between atoms in state $|i\rangle$ and state $|j\rangle$.

When the quantum statistics have no effect on the dynamics of the mean field, we can use the Gross-Pitaevskii (GP) equation to describe the atom laser. The coupled GP equations arising from Eqs.\ (\ref{eqGeneralHamiltonian}) -- (\ref{eqHcoupling}) are given by
\begin{eqnarray}
i \hbar \frac{\partial \psi_t}{\partial t} &=& \bigg( \frac{-\hbar^2}{2m} \nabla^2 + \frac{1}{2} m \omega_t^2 r^2 + U_{tt}|\psi_t|^2 \nonumber \\
&&  \hspace{0.5cm} + U_{tu} |\psi_u|^2 \bigg) \psi_t- \hbar \Omega ({\mathbf{r}})  \psi_u \label{eqGPEOMpsitrapped} \\
i \hbar \frac{\partial \psi_u}{\partial t} &=& \bigg( \frac{-\hbar^2}{2m} \nabla^2  - \delta + U_{uu}|\psi_u|^2  \nonumber \\
&& \hspace{0.5cm} + U_{tu} |\psi_t|^2 \bigg) \psi_u - \hbar \Omega^* ({\mathbf{r}}) \psi_t.  \label{eqGPEOMpsiuntrapped}
\end{eqnarray}
These equations can be solved numerically in one, two or even three dimensions, depending on the spatial resolution required and computational resources available.

As we have included position dependence in the matter fields, the effective Hamiltonian describes the full multimode nature of the problem, and also includes non-Markovian effects.

\section{Linewidth calculations} \label{secLinewidth}
Unlike the optical laser, atomic fields do not have a simple proportional relationship between the energy and momentum spectra, and hence the linewidth in one does not translate trivially to a linewidth in the other.  In free space, the distinction is largely irrelevant as both spectra are static, but this is not always true.  In most current experimental atom lasers, atoms are outcoupled from a trap and allowed to fall under gravity. As they fall, they gain kinetic energy at the expense of potential energy. This results in a kinematic compression effect, leading to a narrower spread in momentum the further the atoms fall, while the energy spread remains constant.  The energy spread of the beam is thus a more stable measurement of the beam's linewidth than the momentum spread. 

Before numerically solving the full equations (\ref{eqGPEOMpsitrapped}) and (\ref{eqGPEOMpsiuntrapped}), it is worthwhile studying a number of simplified versions of the problem in order to extract as much analytic insight as possible from the problem. To this end, we will first examine the problem in the case where the atomic nonlinearities in the condensate are negligible, and assume the condensate remains single mode. We then examine the case where the condensate is allowed to be multimode, but still linear. Finally we consider the fully general case numerically. 

\subsection{Single-mode condensate with no nonlinearity}
\label{secSingleModeLinear}
To begin, we assume the condensate begins in a single mode and remains single mode due to the outcoupling being weak enough such that there is negligible back action on the condensate. This is the simplest possible case, and will result in the minimum possible linewidth achievable in a non-pumped atom laser.

We will also assume that the condensate has no nonlinear interactions, which means we set $U_{tt} = U_{uu} = U_{tu} = 0$ in Eqs.\,(\ref{eqGPEOMpsitrapped}) and (\ref{eqGPEOMpsiuntrapped}). This represents a regime where the condensate contains few atoms, the trapping potential is weak, or the \emph{s}-wave scattering length of the atomic species is small.

\begin{figure}[htb]
\begin{center}
\includegraphics[width=7.5cm,height=6cm]{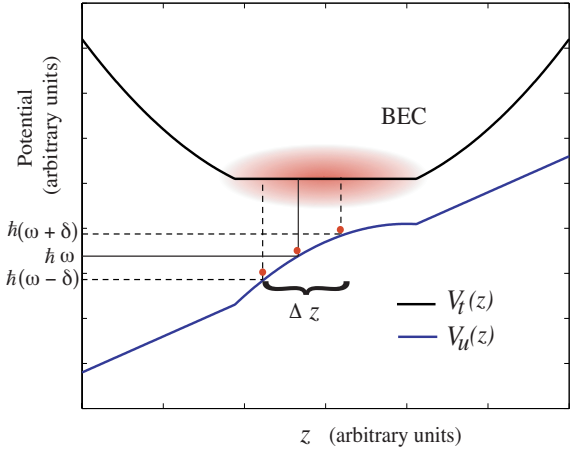}
\caption{Naive description of the energy spread of outcoupled atoms due to a ``balls on a hill'' model. The atomic transition has a resonant width $\delta$ such that the coupling happens over a region $\Delta z$, corresponding to a change in potential of $\hbar\delta$. This leads to a spread in energies of $\hbar\delta$ for the outcoupled atoms. This model neglects the wave-like nature of the atoms, and that there is interference between the different energies. }
\label{fig:balls_on_a_hill}
\end{center}
\end{figure}

In this situation one might naively assume that the linewidth of the atom laser is the power-broadened linewidth of the atomic transition. To see this, one can consider the outcoupling process as a classical picture of ``atoms rolling down a hill", where the untrapped atoms are created on a potential hill arising from the fact that the potentials the trapped and untrapped atoms see are different. In this naive picture the atoms appear on the potential hill with a spread in position given by setting the energy spread equal to the power broadened linewidth, as depicted in Figure~\ref{fig:balls_on_a_hill}. This picture is false, however, as the wave-like nature of the atoms can cause destructive interference of some energies and constructive interference of other energies. It is necessary to take the wave-like nature of the atoms into account to accurately describe the energy spread.  

The Hamiltonian given by Eq.\,(\ref{eqGeneralHamiltonian}) consists of terms describing the trapped matter field, the untrapped matter field, and the coupling between the two. We denote the ground state energy eigenfunction of the trapping Hamiltonian $\hat{H}_{\mathrm{trap}}$ as $\phi_{t}(x)$ and the energy eigenfunctions of the beam Hamiltonian $\hat{H}_{\mathrm{beam}}$ as $\phi_u(q,x)$, where $q$ is any convenient continuous parameter that can label the energies of the atom laser beam. The eigenvalues of $\phi_{t}(x)$ and $\phi_u(q,x)$ are $\hbar \omega_{0} = \hbar \omega_t /2$ and $\hbar \omega(q)$ respectively.

We can now expand the wave functions of the trapped and untrapped matter fields as
\begin{eqnarray}
\psi_t(x,t) &=& \alpha_0(t)\phi_{t}(x) \\
\psi_u(x,t) &=& \int_{-\infty}^{\infty} \beta(q,t) \phi_u(q,x) \,dq.
\end{eqnarray}
Under the approximations described above, Eqs.\,(\ref{eqGPEOMpsitrapped}) and (\ref{eqGPEOMpsiuntrapped}) become
\begin{eqnarray}
i\dot{\alpha}_0 &=& \omega_{0}\alpha_0 - \Omega\int_{-\infty}^{\infty} A(q) \beta(q,t) \,dq \label{alphadot1} \\
i\dot{\beta}(q,t) &=& (\omega(q)-\delta)\beta(q,t) - \Omega^{*}A^{*}(q) \alpha_0(t) \label{betadot1} ,
\end{eqnarray}
where $A(q) = \int_{-\infty}^{\infty} \phi_{t}^*(x)\phi_u(q,x)\Lambda(x)\,dx$, and $\Lambda(x) = \Omega(x)/\Omega$ represents the spatially-dependent part of the electromagnetic field coupling the atoms out of the trap. In the case of Raman outcoupling $\Lambda(x) = e^{i k_0 x}$, and in the case of rf outcoupling $\Lambda(x) = 1$.

Eqs.\,(\ref{alphadot1}) and (\ref{betadot1}) can easily be solved numerically, but we first derive an approximate analytic solution to gain insight into how the linewidth scales with various parameters.

By making the transformation
\begin{eqnarray}
\tilde{\alpha}_0(t) &=& \alpha_0(t)e^{i\omega_{0} t}  \\
\tilde{\beta}(q,t) &=& \beta(q,t)e^{i(\omega(q) -\delta)t} , 
\end{eqnarray}
we obtain
\begin{eqnarray}
i\dot{\tilde{\alpha}}_0 &=&  - \Omega\int_{-\infty}^{\infty} A(q) \tilde{\beta}(q,t)e^{i\Delta\omega(q)t} \,dq \label{alphadot2} \\
i\dot{\tilde{\beta}}(q,t) &=&  - \Omega^{*}A^{*}(q) \tilde{\alpha}_0(t)e^{-i\Delta\omega(q)t}, \label{eqbetadot2}
\end{eqnarray}
where $\Delta\omega(q) = \omega_{0} - (\omega(q)-\delta)$. Formally integrating Eq. (\ref{eqbetadot2}) and assuming the initial state of the output field is vacuum we obtain
\begin{equation}
\tilde{\beta}(q,t) = i\Omega^{*}A^{*}(q)\int_0^{t} \tilde{\alpha}_0(t^{\prime}) e^{-i\Delta\omega(q)t^{\prime}} \,dt^{\prime}. \label{betasol1}
\end{equation}
Substituting this result into Eq.\,(\ref{alphadot2}) gives
\begin{equation}
\dot{\tilde{\alpha}}_0 = -|\Omega|^2\int_{0}^{t}\int_{-\infty}^{\infty} |A(q)|^2  \tilde{\alpha}_0(t^{\prime}) e^{i\Delta\omega(q)(t -t^{\prime})} \frac{d q}{d \omega} \,d\omega \,dt^{\prime}. 
\end{equation}

To proceed we make use of the fact that in the weak outcoupling regime the momentum spread of the output is much narrower than the momentum spread of the condensate. This means that due to momentum conservation we are only selecting atoms with a narrow range of momenta from the condensate, and thus can assume that the form of $A(q)$ is flat over this range, allowing us to replace $A(q)$ with $A(q_0)$ and take it outside the integral.

Similarly, due to energy conservation, the energy of the outcoupled atoms will be centered around $\hbar(\omega_{0} + \delta)$, meaning most of the dynamics will occur at frequencies close to $\omega_{0} + \delta$. Provided $d q / d \omega(q)$ is slowly varying close to $\omega_{0} + \delta$, it is a valid approximation to replace $d q / d \omega$ with  $ d q / d \omega |_{\omega_{0} + \delta}$. For free space $d q / d\omega \propto \omega^{-1/2}$, which means that for a Raman transition with a large momentum kick it will be  approximately constant. In the case of a gravitational potential, $d q / d \omega$ is constant, so the approximation is exact. Using these approximations we obtain
\begin{eqnarray}
\dot{\tilde{\alpha}}_0 &=& -|\Omega|^2|A(q_0)|^2\left. \,\frac{d q}{d \omega}\,\right|_{(\omega_{0t} + \delta)} \nonumber \\
&& \hspace{1cm} \times \int_{0}^{t} \int_{\omega = -\infty}^{\omega=\infty}  \tilde{\alpha}_0(t^{\prime}) e^{i\Delta\omega(q)(t -t^{\prime})} \,d\omega \,dt^{\prime} \nonumber \\
&=& -\frac{\gamma}{2}\tilde{\alpha}_0(t) , \label{alphadot4} 
\end{eqnarray}
where
\begin{equation}
\gamma = 2\pi|\Omega|^2|A(q_0)|^2\left. \frac{d q}{d \omega}\right|_{(\omega_{0} + \delta)}. \label{eqLinewidthAnalyticSingleMode}
\end{equation}
As $\tilde{\alpha}_0(t=0) =\sqrt{N_0}$, where $N_0$ is the number of atoms in the condensate at $t=0$, the solution to Eq.\,(\ref{alphadot4}) is
\begin{equation}
\tilde{\alpha}_0(t) = \sqrt{N_0}e^{-\frac{\gamma}{2}t} , \label{alphasol}
\end{equation}
Consequently, the condensate number $N(t) = |\alpha_0(t)|^2$ will decay exponentially according to $N(t) = N_0 e^{-\gamma t}$. 

\begin{figure}[htb]
\begin{center}
\includegraphics[width=8cm,height=6cm]{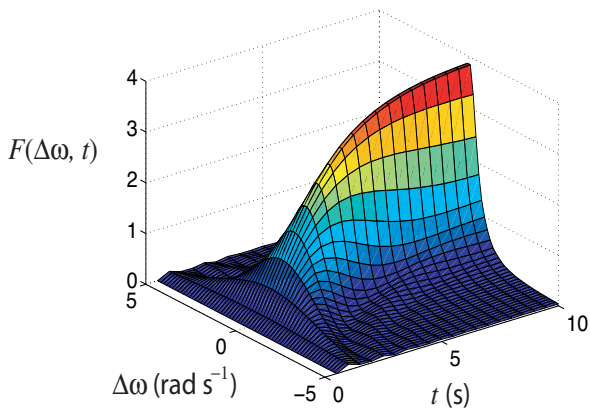}
\caption{$F(\Delta\omega,t)$ for $\gamma = 1$ Hz. As $t \to \infty$, $F(\Delta\omega)$ asymptotically approaches a Lorentzian of FWHM $\gamma$.}
\label{fig_nose}
\end{center}
\end{figure}

The spectrum of the output (i.e. number of atoms per mode $q$) can now be obtained from Eq.\,(\ref{betasol1}). We find
\begin{equation}
|\beta(q,t)|^2 = |\Omega|^2|A(q)|^2 N_0 F(\Delta \omega ,t)
\label{eqSpectrumSingleModeLinear}
\end{equation}
where
\begin{equation}
F(\Delta \omega ,t) = \left(\frac{1 - 2\cos(\Delta\omega t) e^{-\frac{\gamma}{2}t} + e^{-\gamma t}}{\frac{\gamma^2}{4} + \Delta\omega^2}\right).
\end{equation}
This spectrum is plotted in Fig.\,\ref{fig_nose}. 

In the weak coupling limit, when $A(q)$ varies with $\omega$ much more slowly than $F(\Delta\omega,t)$, as $t \rightarrow \infty$ the spectral power density of the output beam $|\beta(q)|^2$ becomes a Lorentzian with a full-width-at-half-maximum given by $\gamma$. Thus, in the long time limit, the fundamental limit to the linewidth of an atom laser is related to the time it takes to drain the condensate: $\gamma = \tau_{drain}^{-1}$, where $\gamma$ is the spectral linewidth (measured in rad s$^{-1}$), and $\tau_{drain}$ is the $1/e$ drain time of the condensate, measured in seconds. This result says that the linewidth of a pulse of atoms coupled out of the condensate sufficiently weakly is only limited by Fourier arguments, although it should be noted that most atom laser experiments are in a stronger coupling regime.  The weak coupling limit gives us the same separation of timescales found in optical cavities, and thus the spectrum is identical to the spectrum of photons draining out of an optical cavity.   As the average flux from our atom laser is $\mathcal{F}_{av} = N_0/ \Delta t$, (where $\Delta t$ is either the drain time of the condensate, or an artificially imposed cut off time of our atom laser pulse), and our spectral linewidth is always limited to $\delta\omega \geq 1/\Delta t$, we obtain the inequality
\begin{equation}
\frac{\mathcal{F}_{av}}{\delta\omega}  \leq N_0 \label{haines_law},
\end{equation}
relating the average flux and spectral linewidth of an unpumped atom laser. 

As a number of approximations were used to obtain Eqs.\,(\ref{eqSpectrumSingleModeLinear}) and (\ref{eqLinewidthAnalyticSingleMode}), we also solved Eqs. (\ref{alphadot1}) and (\ref{betadot1}) numerically and compared the results. For our numerical model we took parameters typical to a $^{87}$Rb atom laser such as one described in \cite{robinsET2006}, and chose $k_0 = 10^{7}\,$m$^{-1}$, $\omega_t = 50\,$rad$\,$s$^{-1}$, and $\Omega = 50\,$rad$\,$s$^{-1}$. We assumed outcoupling into free space, and with the free space dispersion relation Eq. (\ref{eqLinewidthAnalyticSingleMode}) becomes
\begin{equation}
\gamma = \sqrt{\pi} |\Omega|^2 \sqrt{\frac{m}{\hbar \omega_t}} \frac{1}{k_0}.
\label{eqLinewidthSingleModeFreeSpace}
\end{equation}

Table \ref{linewidthtable} shows the comparison between the analytic theory and the numerical simulations. Overall, there is good agreement, allowing us to use Eqs. (\ref{eqSpectrumSingleModeLinear}) and (\ref{eqLinewidthSingleModeFreeSpace}) with some confidence.
\begin{table}
 \begin{center}
  \begin{tabular}{| l | l c c c |}
  	\hline
$k_0$ (m$^{-1}$) & $\Omega$ (rad s$^{-1}$) & $1/\tau$ (Hz) & $\gamma_n$ (rad s$^{-1}$) & $\gamma_a$ (rad s$^{-1}$) \\ \hline \hline

			& 400* 	&   270*	       & 320*	 & 292 *	 \\ 
$1 \times 10^7$	& 100	&   18	 	& 18      & 18.3 	 \\  
			& 25	&  1.1		& 1.2      & 1.14    \\ 
 			& 10	& 0.18		& 0.18     & 0.183 \\ \hline
	
			& 100 	&   35	       & 37	 & 36.5   \\ 
$5 \times 10^6$	& 25	&   2.3	 	& 2.4      & 2.28 	 \\  
 			& 10	& 0.36		& 0.35     & 0.365 \\ \hline
			
			&100* 	&   100*	       & 245*	 & 182*   \\ 
$1\times10^6$ 		& 25	&   11	 	& 13      & 11.4 	 \\  
 			& 10	& 1.8		& 1.8     & 1.83     \\ \hline
	
	\end{tabular}
	\caption{Comparison of condensate drain time $1 / \tau$ and the long-time linewidth of an atom laser for different values of $\Omega$, and $k_0$. $\gamma_n$ and $\gamma_a$ represent the linewidths for the analytic theory and numerical simulation respectively. The table shows close agreement between our approximate analytic result and our numeric calculation. The entries marked with an asterisk display poor agreement between the analytic and numeric results. This is because the coupling is sufficiently large such that the approximation made in Eq.\,(\ref{alphadot4}) is invalid.}
\label{linewidthtable}
  \end{center}
\end{table}

The key point of this semiclassical analysis is that for an ideal, single-mode, non-pumped atom laser, the linewidth is given by the inverse of the drain time of the condensate. Consequently it can be made as narrow as desired by reducing the outcoupling strength arbitrarily. The trade off is that this arbitrarily narrow linewidth comes at the expense of reduced flux.

\subsection{Multimode condensate with no nonlinearity}

We now consider a more realistic model, where the condensate is not constrained to remain in a single mode, although we still assume the nonlinearities are negligible.

We proceed as in the previous section, except now we allow the condensate to be multimode, and for simplicity treat both the condensate modes and atom laser modes as discrete. The wave functions of the trapped and untrapped atoms are now expanded as
\begin{eqnarray}
\psi_t(x,t) &=& \sum_{n} \alpha_n(t) \phi_{tn}(x) \label{linewidth_modebasis1} \\
\psi_u(x,t) &=& \sum_n \beta_n(t) \phi_{un}(x) \label{linewidth_modebasis2}
\end{eqnarray}
where $\phi_{tn}(x)$ and $\phi_{un}(x)$ are the $n$th eigenstates of the Hamiltonians $\hat{H}_{\mathrm{trap}}$ and $\hat{H}_{\mathrm{beam}}$ respectively. We denote their eigenvalues by $\hbar \omega_{tn}$ and $\hbar \omega_{un}$. Proceeding as before, we find that the equations of motion in an appropriate rotating frame are given by
\begin{eqnarray}
i\dot{\tilde{\alpha}}_m &=& -\Omega\sum_n A_{mn}\tilde{\beta}_n e^{i(\omega_{tm} - (\omega_{un}-\delta))t} \label{alphadot_twidle} \\
i\dot{\tilde{\beta}}_m &=& \Omega^{*}\sum_n A^{*}_{mn}\tilde{\alpha}_n e^{-i(\omega_{tn} - (\omega_{um}-\delta))t} .  \label{betadot_twiddle}
\end{eqnarray}
where $A_{nm} = \int_{-\infty}^{\infty} \phi^{*}_{tm}(x)\phi_{un}(x) \Lambda(x)\,dx$.

When the coupling $\Omega$ is weak, the phase rotation of $\tilde{\alpha}_m(t)$ and $\tilde{\beta}_m(t)$ is approximately at zero frequency. This means that over long times the only significant contribution to the growth of $\tilde{\beta}_m(t)$ is from the trapped mode $\tilde{\alpha}_{n}$ with frequency
\begin{equation}
\omega_{tn} = \omega_{um}-\delta,
\label{eqEnergyConservationMultimode}
\end{equation}
as all other modes will on average cause no net growth on time-scales much larger than $\tau = (\omega_{tn} - (\omega_{um}-\delta))^{-1}$.

In the case of rf outcoupling, the momentum kick to the outcoupled atoms is negligible, so the atoms retain the momentum they had when they were in the trap. This means the energy conservation relation (\ref{eqEnergyConservationMultimode}) is the only condition that must be satisfied when considering the output spectrum, and consequently the energy spread of the output is now related to the energy spread of the condensate via the magnitude of the matrix elements $A_{nm}$. Thus the spectrum of the atom laser will essentially mimic the spectrum of the condensate, with the energy peaks in the output beam at frequencies $\omega_{tn} + \delta$ corresponding to different energies in the condensate, moderated by the magnitude of $A_{nm}$, with each peak broadened such that it is the Fourier limit of the outcoupling time, as discussed in the previous section. It is therefore clear that in the rf case any dynamic fluctuations in the BEC will result in an atom laser with a broader spectral linewidth than if the BEC were single mode. 

\begin{figure}[htb]
\begin{center}
\includegraphics[width=8cm,height=7cm]{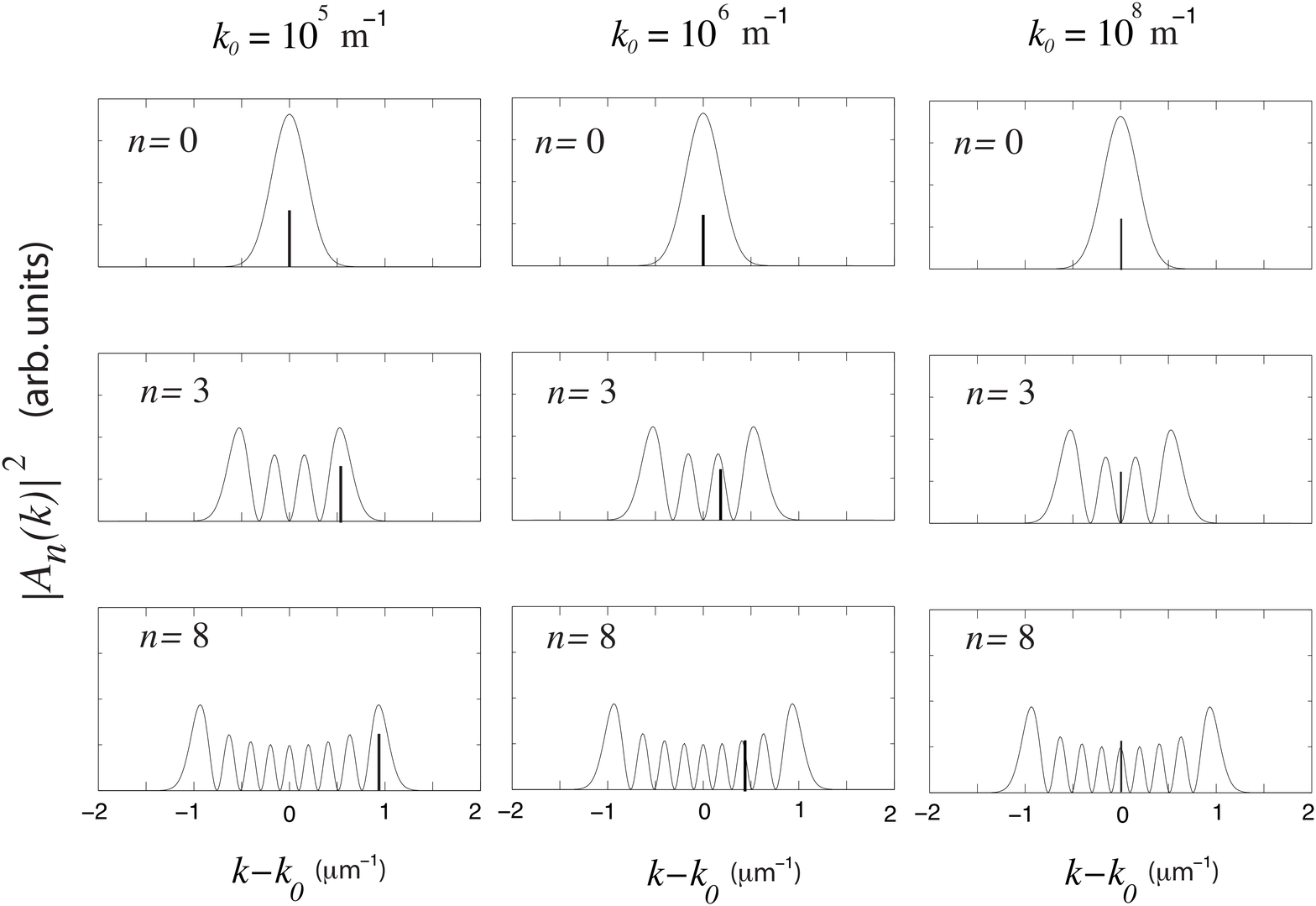}
\caption{$|A_n(k)|^2$ for $n=0$ (top),  $n=3$ (middle) and  $n=8$ (bottom). $k_n = \sqrt{(\delta +\omega_n)2m / \hbar} - k_0$, the place in the $k$-space wave function from where resonant outcoupling occurs, is indicated by a vertical black bar in each case. For $k_0 = 10^5\,$m$^{-1}$, $k_n$ follows the largest `lobe' of $A_n(k)$ for increasingly excited states. For $k_0 = 10^8\,$m$^{-1}$, $k_n$ remains approximately in the center of $A_n(k)$. $\delta = \hbar k_0^2 /2m - \omega_{t0}$ was chosen such that the outcoupling was perfectly on resonance for the zero momentum component of the ground state.}
\label{raman_mode_fig_1}
\end{center}
\end{figure}

The situation is more complicated in the case of a Raman outcoupling scheme with a large momentum kick. If we assume that our condensate is initially in the superposition 
\begin{equation}
\psi_t(x) = \sum_n \alpha_n \phi_{tn}(x) ,
\end{equation}
then, for long times, only the states that satisfy the energy resonance $\omega_{tn} = \omega_{uj} - \delta$ will be present in the output, just as in the rf case. However, now there is an additional constraint arising from the dispersive nature of the atoms. Taking free space as an example, we have 
\begin{equation}
\omega_{uj} = \frac{\hbar k_j^2}{2m} , \label{raman_energy_conserve2}
\end{equation}
where $\hbar k_j$ is the momentum of $\phi_{uj}(x)$. Thus the output will only contain momentum states 
\begin{equation}
k_{j} = \sqrt{\frac{2m(\delta +\omega_n)}{\hbar}}.
\end{equation}
However, conservation of momentum demands $k_j = k+k_0$, where $k$ is the initial momentum of the atoms. The source of outcoupled atoms will then be the component of the condensate with momentum 
\begin{equation}
k_n =  \sqrt{\frac{2m(\delta +\omega_n)}{\hbar}} - k_0.
\end{equation}
The relative amplitude of the component of the output with momentum $k_j$ is then proportional to the value of $A_n(k)$, the $k$-space representation of $\phi_{t,n}(x)$, evaluated at $k_n$. This can lead to interesting effects. As an example, we choose our trapped system as a harmonic oscillator. When $k_0$ is small, the spacing between the output $k$ modes is of order $\Delta k_j \sim \sqrt{2 m \omega_t/\hbar}$, which is about the spacing of the `lobes' in the $k$ space representation of $\phi_{tn}(x)$. When $k_0$ is large, $\Delta k_j \sim 0$, so the outcoupling always happens close to the center of the momentum space wave function. As $A_n(k)$ are Hermite Gaussians in the case of the harmonic oscillator, $A_n(k=0) = 0$ for $n$ odd, so there is no outcoupling from odd modes. This is represented in Fig.\,\ref{raman_mode_fig_1}.

\begin{figure}[htb]
\begin{center}
\includegraphics[width=8cm,height=7cm]{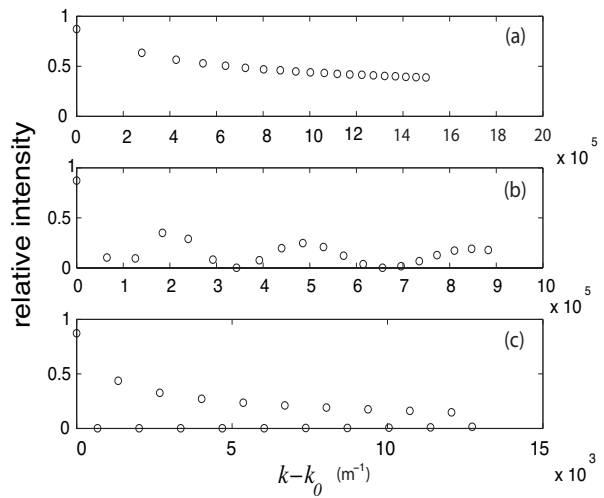}
\caption{The relative intensities of the momentum components of the atom laser beam corresponding to the first 20 condensate eigenmodes for (a) $k_0 = 10^{5}\,$m$^{-1}$ (b) $k_0 = 10^{6}\,$m$^{-1}$ and (c) $k_0 = 10^{8}\,$m$^{-1}$. For small $k_0$ ($k_0 = 10^{5}\,$m$^{-1}$), the relative intensity of each mode decreases due to the spreading out of $A_n(k)$. For large $k_0$ ($k_0 = 10^{8}\,$m$^{-1}$), only even modes are present in the output. This is due to $k_n$ falling at nodes of $A_n(k)$ for the odd modes. In the intermediate case, there is complicated structure, with some odd modes and some even modes being attenuated.}
\label{raman_HO_rel_intens}
\end{center}
\end{figure}

The relative intensity of each $k$ component in the output is proportional to $|\alpha_n|^2 |A_n(k_n)|^2$. Fig.\,\ref{raman_HO_rel_intens} shows $|A_n(k_n)|^2$ for different values of $k_0$. 
Fig.\,\ref{raman_HO_rel_intens} was checked against a multimode Gross-Pitaevskii simulation for the first five modes. Close agreement was found for the relative amplitudes, although the momentum resolution was insufficient to accurately resolve the difference in momentum for each peak. For small $k_0$, all modes are present in the output, with the relative intensity of each mode decreasing due to the spreading out of $A_n(k)$. For large $k_0$, only even modes are present in the output. This is due to $k_n$ falling at nodes of $A_n(k)$ for the odd modes. In the cross-over regime (when $k_0$ is of order $\sqrt{m \omega_t / \hbar}$), there is complicated structure in the output with some even and some odd modes severely attenuated in the output. For $^{87}$Rb, at a typical trapping frequency ($\omega_t = 50\,$rad$\,$s$^{-1}$), this cross over occurs at around $k_0 \approx 4\times 10^{5}$ m$^{-1}$, which is much less than the maximum recoil of $k \approx 1.6\times 10^{7}\,$m$^{-1}$ achievable with a two-photon transition, using light of wavelength $\lambda = 780\,$nm. This suggests that this effect should be observable in experiments, although our theory has neglected the atomic interactions, which will complicate the effect.

\subsection{Multimode condensate with nonlinear interactions}

We now turn to an analysis of the complete problem and allow the nonlinear interactions in the condensate to be significant, which is the case in many experimentally realizable atom lasers.

The existence of nonlinearities makes a difference to a number of properties of the condensate and the atom laser, affecting things such as mode shapes, memory functions and classical density fluctuations. The change that is most relevant to the linewidth, however, is the fact that the energy of the condensate is now dependent on the number of atoms in the condensate.

To understand this, we note that when an atom in the condensate is flipped from a trapped to an untrapped state, it experiences a mean field potential that depends on the density distribution of the condensate. As the density of the untrapped field $|\psi_u|^2$ is much less than that of the trapped field inside the condensate, it is clear from Eq. (\ref{eqGPEOMpsiuntrapped}) that this mean field potential is given by
\begin{equation}
V_{\mathrm{mf}} = U_{uu} |\psi_t({\mathbf{r}})|^2,
\label{eqMeanFieldForUntrappedAtoms}
\end{equation}
where $U_{uu}$ is defined by Eq.\,(\ref{eqNonlinearPotentialDefinition}). As density is always positive and increases towards the center of the condensate, this results in a repulsive force on the untrapped atoms that accelerates them out of the BEC.

As an example, we consider the Thomas-Fermi limit, where the nonlinear energy of the condensate is considerably larger than the kinetic energy and the wave function for the condensate can be found analytically. Assuming a harmonic trapping potential with frequency $\omega_t$, the atomic density in the condensate is given by
\begin{equation}
|\psi_t({\mathbf{r}})|^2 = \frac{1}{U_{tt}} \left( \mu(N) - \frac{1}{2} m \omega_t^2 r^2 - m g z \right)
\label{eqThomasFermiDensity}
\end{equation}
where 
\begin{equation}
\mu(N) = \frac{m \omega_t^2}{2} \left(\frac{15 N U_{tt}} {4 \pi m \omega_t^2} \right)^{2/5}. \label{eqmu3D} 
\end{equation}
Eqs.\,(\ref{eqMeanFieldForUntrappedAtoms}) and (\ref{eqThomasFermiDensity}) show that after the atoms have been flipped into an untrapped state, they slide down a quadratic potential hill of height $\mu$, giving them a kinetic energy $\mu$ as they leave the condensate.

Assuming the condensate is not pumped, its atom number will inevitably reduce during the outcoupling process. As the number of atoms $N$ in the condensate falls, the the chemical potential also falls, resulting in a situation where atoms outcoupled later in time will have a smaller kinetic energy as they leave the condensate compared to atoms outcoupled at an earlier time. Consequently, the atom laser beam will consist of atoms with a wide spread of energies; a spread that can be as large as $\mu(N_0)$ if all the atoms are outcoupled.

To demonstrate this effect we numerically solve the full Eqs.\,(\ref{eqGPEOMpsitrapped}) and (\ref{eqGPEOMpsiuntrapped}) for a situation with experimentally realistic parameters and a large nonlinearity. The equations were solved in one dimension only, using a dimensional reduction procedure where the nonlinear potentials $U_{ij}$ were scaled by a transverse area corresponding to the cross sectional area of the beam \cite{steelET1998}. This not only makes the computation far more tractable, but also removes additional complications that obscure the linewidth such as part of the mean-field kick being transferred into transverse modes of the laser. The simulation couples into free space rather than a gravitational potential, meaning the momentum space wave function of the beam can be used to give the linewidth.

\begin{figure}[htb]
\begin{center}
\includegraphics[width=8cm,height=7cm]{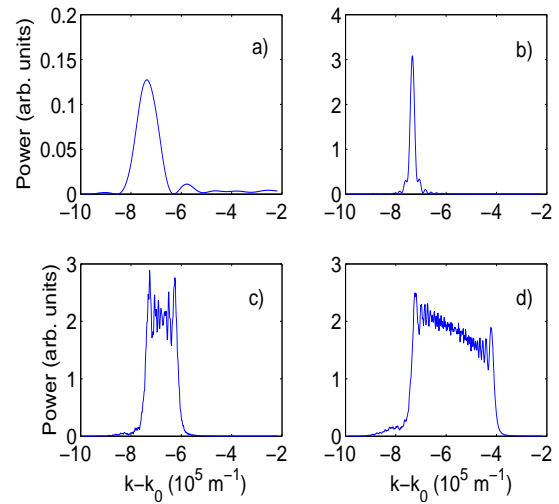}
\caption{Momentum space density of an atom laser beam after a) 20\,ms, b) 100\,ms, c) 400\,ms, and d) 1000\,ms of outcoupling. After 1000ms 65\% of the atoms in the condensate have been outcoupled. Parameters: $N_0 = 10^{6}$, $\omega = 150\,$rad$\,$s$^{-1}$, $a=4\times 10^{-11}\,$m, $k_0 = 3.2\times 10^{6}\,$m$^{-1}$.}
\label{figLineBroadeningFromChirp}
\end{center}
\end{figure}

Figure~\ref{figLineBroadeningFromChirp} shows the result of the simulation, displaying snapshots of the beam's momentum space wave function at various points in time. Initially the momentum is centered at $k_{\mathrm{cent}}=\sqrt{k_0^2 + 2 m \mu(N_0) / \hbar^2}$, and the linewidth (i.e. momentum spread of the beam) begins to narrow in accordance with the Fourier argument laid out in Section \ref{secSingleModeLinear}.  After enough atoms have been outcoupled to significantly change the chemical potential, however, new atoms appearing in the beam have lower and lower energies, resulting in the line center being ``chirped''. Consequently the effective linewidth becomes ever broader, ultimately spanning all momentum states between $k_0$ and $k_{\mathrm{cent}}$ when the entire BEC has been drained.

The complicated structure seen in Figure~\ref{figLineBroadeningFromChirp} arises from the fact that an atom laser beam is a complex field, and as the frequency of condensate phase evolution is changing, there can be destructive interference between atoms emitted with a particular energy and atoms emitted at a later time with the same energy but out of phase.

\section{Methods to reduce linewidth}\label{secLinewidthReduction}

In this section we review possible methods for reducing the linewidth from an unpumped atom laser in various parameter regimes.

\subsection{Weak outcoupling}
\label{secWeakOutcoupling}

The most obvious way to minimize the linewidth of an atom laser is to outcouple extremely weakly, as this increases the drain time, effectively without limit. Thus, provided the condensate nonlinearities are negligible, the Fourier arguments in Section \ref{secSingleModeLinear} demonstrate that the linewidth can be made arbitrarily narrow.

In the case where condensate nonlinearities are \emph{not} negligible, weak outcoupling still succeeds in reducing the linewidth as the condensate will undergo almost no depletion, meaning the chemical potential is static and there is no chirp of the line center. However, unlike the case where nonlinearities can be ignored, we cannot weakly outcouple all the atoms --- we must ensure that over the entire duration of the experiment the change in chemical potential is less than the minimum linewidth we are willing to accept. In the case of a strongly nonlinear condensate, if we require the temporal linewidth to be less than $\delta \omega$, then by Eq.\,(\ref{eqmu3D}) we must ensure that the number of atoms outcoupled from the condensate is less than
\begin{equation}
\Delta N < \frac{5}{m \omega_t^2 \hbar} \left( \frac{4 \pi m \omega_t^2}{15 U_{tt}} \right)^{2/5} N_0^{3/5} \, \delta \omega
\end{equation}
where $\omega_t$ is the harmonic trapping frequency and $N_0$ is the number of atoms initially in the condensate.

\begin{figure}[htb]
\begin{center}
\includegraphics[width=7cm,height=6cm]{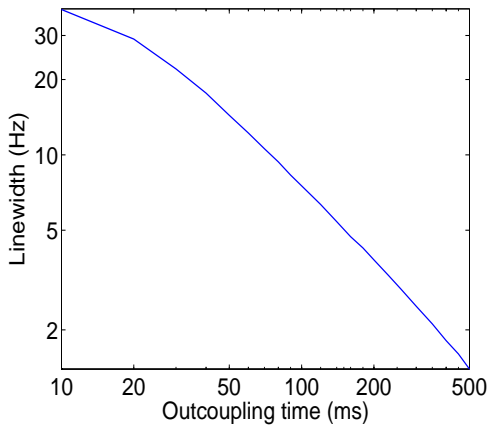}
\caption{Linewidth narrowing as a function of outcoupling time. Parameters: $N=10^{7}$, $\omega_t = 250\,$rad$\,$s$^{-1}$, $a=3\times 10^{-9}\,$m, $k_0 = 10^{7}\,$m$^{-1}$.} 
\label{figLinewidthNarrowing_a3nm_loglog}
\end{center}
\end{figure}

This approach can be arbitrarily effective if high flux is not important.   We solved Eqs. (\ref{eqGPEOMpsitrapped}) and (\ref{eqGPEOMpsiuntrapped}) for a highly nonlinear system, and examined the linewidth of the beam over time. Over the simulation approximately 10 atoms were removed from the condensate, corresponding to extremely weak outcoupling. The results are shown in Figure\,\ref{figLinewidthNarrowing_a3nm_loglog}. In the long time limit the curve shown in Figure \ref{figLinewidthNarrowing_a3nm_loglog} is linear with a slope of -1, indicating that linewidth is inversely proportional to the outcoupling time. This agrees with the Fourier arguments our approximate single-mode linear theory predicts.

The difficulty with weak outcoupling is that high flux is one of the more desirable qualities in a laser.  What have demonstrated here is that in the ultra-low flux limit the atom laser can be regarded in some sense as having a very narrow linewidth with a slowly moving line center.  In practice, this flux limit will make any experiments impractical, so we now consider a method to achieve narrow linewidth without sacrificing flux.

\subsection{Chirp compensation}

The source of the drift of the line center is the mean field potential that untrapped atoms experience as they leave the condensate. If we ignore gravity, this mean field potential can be found from Eqs. (\ref{eqMeanFieldForUntrappedAtoms}) and (\ref{eqThomasFermiDensity}) and is given by
\begin{equation}
V_{\mathrm{mf}} = \mu(t) - \frac{1}{2} m \omega_t^2 r^2,
\end{equation}
where we have taken $U_{tt} = U_{uu}$ and allowed the chemical potential time dependence to take into account condensate depletion. If we choose the outcoupling point to be the center of the condensate, atoms acquire the full chemical potential $\mu(t)$ worth of energy on their way out of the condensate. However, as the condensate depletes, $\mu(t)$ decreases, meaning atoms outcoupled later have less energy, broadening the linewidth.

The solution is to begin outcoupling from a point away from the center of the condensate, so that initially atoms do not acquire the full $\mu(t)$ worth of energy as they slide down the potential hill. If we then move the outcoupling point back in towards the center of the condensate as it depletes, it is possible to ensure that atoms outcoupled later acquire the same amount of energy as those outcoupled earlier, thus removing the chirp effect.

This shifting of the outcoupling point can be accomplished by making the two-photon detuning $\delta$ time dependent. If we wish to initially begin outcoupling atoms from a distance $r_0$ from the minimum of the magnetic trap (which will coincide with the center of the condensate if gravity is ignored), and sweep this point towards the center of the trap in such a way that atoms always leave the condensate with the same energy, then, since the energy of atoms as they leave the condensate is
\begin{equation}
E_{\mathrm{out}} = \delta(t) + \mu(t)
\end{equation}
we need to choose
\begin{equation}
\delta(t) = \frac{\hbar^2 k_0^2}{2 m} - \mu(t) + \mu(0) - \frac{1}{2} m \omega_t^2 r_0^2.
\label{eqTimeDependentDetuning}
\end{equation}

\begin{figure}[htb]
\begin{center}
\includegraphics[width=7cm,height=6cm]{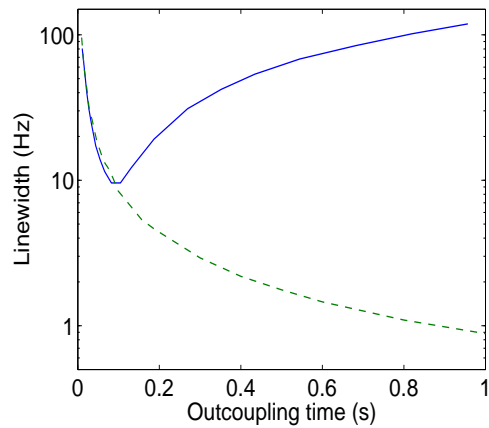}
\caption{Comparison of linewidth over time for a condensate with significant nonlinearities that becomes depleted. Solid line shows standard outcoupling; dashed line shows the effect of our chirp correction scheme. Parameters: $N_0 = 10^{6}$, $\omega_t = 150\,$rad$\,$s$^{-1}$, $a=4\times 10^{-11}\,$m, $k_0 = 3.2\times 10^{6}\,$m$^{-1}$.} 
\label{figLinewidthOverTimeNoChirpCorrection}
\end{center}
\end{figure}

To test this scheme we numerically solved Eqs.\ (\ref{eqGPEOMpsitrapped}) and (\ref{eqGPEOMpsiuntrapped}) for the same nonlinear system that was considered in Section \ref{secWeakOutcoupling}, carrying out simulations with both a fixed two-photon detuning and a time-dependent two-photon detuning given by Eq.\,(\ref{eqTimeDependentDetuning}). The behavior of the atom laser linewidth over time is shown in Figure~\ref{figLinewidthOverTimeNoChirpCorrection}. It is clear that using a time-dependent detuning prevents the deleterious chirp of the line center, and recovers the underlying narrow spectrum of the laser output.

\subsection{Pumping}

A pumped atom laser operating at steady state would experience no chirp, as the energy of the lasing mode would be stable by definition.  Lasers operating well over threshold also experience mode-selection effects that help provide a stable mode, although previous work suggests that it may be nontrivial to operate in this regime \cite{haineET2003,johnssonET2005}.  Pumping can also induce gain-narrowing to combat the linewidth-broadening effects of quantum noise.  A continuous pumping scheme for BECs has not been demonstrated, however, and consideration of quantum noise contributions to the linewidth of an atom laser requires a model that goes beyond the semi-classical approximation.  Examination of these quantum effects in zero-dimensional models has occurred in a variety of contexts \cite{graham1998,wisemanET2001,bradleyET2003}, but no model has examined the competition between the multimode effects and the quantum noise in these devices.  A multimode quantum model to investigate the quantum noise contribution to the linewidth of an unpumped atom laser will be the subject of a forthcoming paper.

\section{Conclusions}

We have examined the linewidth of experimentally realistic non-pumped atom lasers in a variety of regimes.  In strong outcoupling regimes, output spectra show a variety of undesirable features.  If the condensate does not remain single mode due to non-Markovian effects causing back action of the beam on the condensate, then the spectrum of the output beam will also be multimode, with peaks corresponding to the energy spacing of the excited modes in the condensate.  We have shown that is possible to use Raman outcoupling and momentum conservation rules to selectively filter out some of the excited modes of the condensate, resulting in a cleaner beam.

In the limit as the coupling strength becomes weaker, we show that the linewidth of non-pumped atom lasers has two main limits.  Weakly outcoupled, single-mode atom lasers with negligible nonlinearities will approach the Fourier limit, where the linewidth is given by the inverse of the out-coupling rate.  This outcoupling rate is a function of the atomic properties as well as the Rabi frequency of the change of state of the atoms.  When the condensate has significant nonlinearities, as is the case is most current experimental atom lasers, depletion of the condensate causes the chemical potential to decrease over time, resulting in a downwards ``chirp'' in the central frequency of the output beam during the outcoupling process. This effect causes significant broadening of the linewidth of the beam.  If high flux is required, and therefore a significant fraction of the condensate must be outcoupled, we demonstrate that it is feasible to use a chirp compensation scheme such as sweeping the detuning of the state-changing outcoupling process.

\acknowledgments
This work was supported by the Centre of Excellence program of the Australian Research Council and the APAC National Supercomputing Facility.


\begin{thebibliography}{0}
\bibitem{kasevich}M. A. Kasevich, Science, {\bf 298} 1363 (2002).

\bibitem{berman}Atom Interferometry, ed. Paul R. Berman (Academic Press, San Diego, 1997).




\bibitem{petersET1999} A. Peters, K. Y. Chung and S. Chu, Nature {\bf 400}, 849 (1999).

\bibitem{gustavsonET2000} T. L. Gustavson, A. Landragin and M. A. Kasevich, Class.\ Quantum Grav.\ {\bf 17}, 2385 (2000).

\bibitem{mcguirkET2002} J. M. McGuirk, G. T. Foster, J. B. Fixler, M. J. Snadden, and M. A. Kasevich, Phys.\ Rev.\ A {\bf 65}, 033608 (2002).

\bibitem{fattoriET2003} M. Fattori, F. Lamporesi, T. Petelski, J. Stuhler, and G. M. Tino, Phys.\ Lett.\ A {\bf 318}, 184 (2003).

\bibitem{wichtET2001} A. Wicht, J. M. Hensley, E. Sarajlic, S. Chu, in \emph{Proc. 6th Symp. Frequency Standards and Metrology}, ed. by P. Gill (World Scientific, Singapore, 2001).

\bibitem{cladeET2006} P. Clad\'{e}, E. de Mirandes, M. Cadoret, S. Guellati-Kh\'{e}lifa, C. Schwob, F. Nez, L. Julien, and F. Biraben, Phys.\ Rev.\ Lett.\ {\bf 96}, 033001 (2006).

\bibitem{toriiET2000} Y. Torii, Y. Suzuki, M. Kozuma, T. Sugiura, T. Kuga, L. Deng and E. W. Hagley,  Phys.\ Rev.\ A {\bf 61}, 041602(R) (2000).

\bibitem{lecoqET2006} Y. Le Coq, J. A. Retter, S. Richard, A. Aspect, and P. Bouyer, Appl.\ Phys.\ B {\bf 84}, 627 (2006).graham1998

\bibitem{reid} M. D. Reid and P. D. Drummond, Phys.\ Rev.\ Lett.\ {\bf 60}, 2731, (1988).

\bibitem{kheruntsyan2005} K. V. Kheruntsyan, M. K. Olsen and P. D. Drummond, Phys.\ Rev.\ Lett.\ {\bf 95}, 150405 (2005).

\bibitem{haine05} S. A. Haine and J. J. Hope, Phys.\ Rev.\ A {\bf 72}, 033601 (2005).

\bibitem{robinsET2006} N. P. Robins, C. Figl, S. A. Haine, A. K. Morrison, M. Jeppesen, J. J. Hope and J. D. Close, Phys.\ Rev.\ Lett.\ {\bf 96}, 140403 (2006).

\bibitem{hagleyET1999} E. W. Hagley,  L. Deng, M. Kozuma, J. Wen, K.
Helmerson, S. Rolston and W. D. Phillips, Science {\bf{283}}, 1706 (1999).

\bibitem{haineET2006} S. A. Haine, M. K. Olsen and J. J. Hope, Phys.\ Rev.\ Lett.\ {\bf{96}}, 133601 (2006).
\bibitem{steelET1998} M. J. Steel, M. K. Olsen, L. I. Plimak, P. D. Drummond, S. M. Tan, M. J. Collett, D. F. Walls, and R. Graham, Phys. Rev. A {\bf 58}, 4824 (1998).

\bibitem{haineET2003} S. A. Haine and J. J. Hope, Phys.\ Rev.\ A {\bf{68}}, 023607 (2003).

\bibitem{johnssonET2005} M. T. Johnsson, S. A. Haine and J. J. Hope, Phys.\ Rev.\ A {\bf{72}}, 053603 (2005).

\bibitem{graham1998} R. Graham, Phys.\ Rev.\ Lett.\ {\bf 81}, 5262 (1998).

\bibitem{wisemanET2001} H. M. Wiseman and L. K. Thomsen, Phys.\ Rev.\ Lett.\ {\bf{86}}, 1143 (2001).

\bibitem{bradleyET2003} A. S. Bradley, J. J. Hope and M. J. Collett Phys.\ Rev.\ A {\bf{68}}, 063611 (2003).
\end{thebibliography}
\end{document}